\documentclass[conference]{IEEEtran}
\usepackage{color,graphicx}
\usepackage{url}
\usepackage{amsmath}
\usepackage{algpseudocode}
\usepackage{algorithmicx}
\usepackage{algorithm}
\usepackage{epstopdf}
\usepackage{epsfig}
\usepackage{cite}
\usepackage{subcaption}

\ifCLASSINFOpdf
\else
\fi

\makeatletter

\newcommand{\Rmnum}[1]{\expandafter\@slowromancap\romannumeral #1@}
\makeatother
\begin{document}
\title{Analysis of CSAT performance in Wi-Fi and LTE-U Coexistence}

\author{Vanlin Sathya$^\dag$, Morteza Mehrnoush$^*$, Monisha Ghosh$^\dag$, and Sumit Roy$^*$
\IEEEauthorblockN{}
\IEEEauthorblockA{$^\dag$University of Chicago, Illinois, USA. $^*$University of Washington, Seattle, USA. \\
{Email: vanlin@uchicago.edu, mortezam@uw.edu, monisha@uchicago.edu, sroy@u.washington.edu.}}
}
\maketitle

\begin{abstract}
In this paper, we study energy-based Carrier Sense Adaptive Transmission (CSAT) for use with LTE-U and investigate the performance in Wi-Fi/LTE-U coexistence using theoretical analysis and experimental verification using NI USRPs. According to the LTE-U forum specification, if an LTE-U base station (BS) finds a vacant channel, it can transmit for up to 20 ms and turn OFF its transmission for only 1 ms, resulting in a maximum duty cycle of 95\%. In a dense deployment of LTE-U and Wi-Fi, it is very likely that a Wi-Fi access point (AP) will wish to use the same channel. It will start transmission by trying to transmit association packets (using carrier sense multiple access with collision avoidance (CSMA/CA)) through the 1 ms LTE-U OFF duration. Since this duration is vey small, it leads to increased association packet drops and thus delays the Wi-Fi association process. Once LTE-U, using CSAT, detects Wi-Fi, it should scale back the duty cycle to 50\%. We demonstrate in this paper, using an experimental platform as well as theoretical analysis, that if LTE-U is using a 95\% duty cycle, energy based CSAT will take a much longer time to scale back the duty cycle due to the beacon drops and delays in the reception. Hence, in order to maintain association fairness with Wi-Fi, we propose that a LTE-U BS should not transmit at maximum duty cycles (95\%), even if the channel is sensed to be vacant.\\

\end{abstract}

\section{Introduction}

As next generation wireless deployments increase in density, two trends are noticeable: one is the increasing number of small-cell deployments in congested areas such as airports, and the second is the increasing using of unlicensed bands for cellular services, \emph{i.e.,} LTE being deployed in unlicensed 5 GHz bands that were used primarily by Wi-Fi. The latter gives rise to coexistence problems between the two systems since LTE and Wi-Fi use very different medium access control (MAC) protocols, with LTE using a scheduled approach and Wi-Fi using carrier sense multiple access with collision avoidance (CSMA/CA).

The issue of Wi-Fi and LTE coexisting on the unlicensed bands at 5 GHz has received increased attention lately, driven in part by the development of two recent standardization efforts: LTE-U~\cite{LTEU}, developed by the LTE-U Forum and LTE-LAA developed by 3GPP~\cite{3GPP}. The two specifications differ in the way coexistence is implemented. LTE-U uses a duty cycling approach combined with Carrier Sense Adaptive Transmission (CSAT) to determine the duty cycle based on Wi-Fi occupancy, while LTE-LAA uses a similar mechanism as Wi-Fi. In this paper, we only consider LTE-U with CSAT, and in particular focus on the association fairness problem~\cite{G_whitepaper} as opposed to the throughput fairness problem which in general has been widely studied~\cite{li2016modeling,chen2016optimizing,cano2016unlicensed,sagari2015coordinated,chen2017embedding}. In our previous work~\cite{WCNC}, we addressed the issue of association fairness when Wi-Fi and LTE-U coexist on the same channel and showed that if LTE-U were transmitting at 20 ms ON and 1 ms OFF, a significant number of W-Fi beacons would either not be transmitted at all or not received at the client, thus leading to delays in association. However, that work did not have a CSAT implementation which would scale back the duty cycle once Wi-Fi was detected. Hence, in this paper we include a CSAT implementation and study a different problem: how long would it take for LTE-U to scale back its duty cycle to 50\% once a Wi-Fi AP started coexisting on the same channel? We show, via experiments and theoretical analysis, that this scale-back time is a function of the initial duty cycle being used by LTE-U: it takes significantly longer to scale back from a 95\% duty cycle than from a 80\% duty cycle since the CSAT algorithm takes longer to detect the presence of Wi-Fi. We only consider energy-based CSAT, i.e. the CSAT algorithm is triggered only by energy levels rather than preamble-based CSAT which would detect the Wi-Fi preambles. 
 
\par The rest of the paper is organized as follows. Section II presents the coexistence system model we consider and the association process in LTE-U Wi-Fi coexistence. Section III describes the motivation for using an initial duty cycle of 80\% instead of 95\%. Section IV describes the proposed energy based CSAT algorithm. Section V presents the calculation for the expected delay in receiving the initial Wi-Fi beacons. Experimental setup and results are presented in Section VI and conclusions in Section VII.

\section{Coexistence System Model}

In this section, we present the system model that we consider in this paper and a brief explanation of the association process in LTE-U Wi-Fi coexistence.
%
% \begin{figure}[htb!]
%\begin{center}
%\includegraphics[totalheight=2.3cm,width=7.3cm]{dc1.eps}
%\end{center}
%\caption{(a) Cell A and Cell B use Wi-Fi on same channel \& Cell C and Cell D use Wi-Fi on different channel, and (b) Cell A switches to LTE-U}
%\label{asso}
%\end{figure}

\subsection{Coexistence System Model}

We consider the situation where both Wi-Fi and LTE-U operate on the same channel in the unlicensed 5 GHz band. The LTE-U transmissions in the unlicensed band are present only on the downlink, with all uplink traffic (acknowledgments and control) being transmitted on a licensed channel. We assume that the LTE-U BS operates in full buffer mode and utilizes the entire bandwidth with a higher modulation scheme (\emph{i.e.,} 64-QAM). We consider both active and passive association, where Wi-Fi clients transmit probe request packets during active scanning and receive beacon packets during passive scanning. Beacon frames are also transmitted with CSMA/CA, i.e. the AP needs to check for the availability of the channel before sending beacon packets. Probe response packets transmitted by the AP (in response to probe requests from Wi-Fi clients) are unicast packets with a corresponding acknowledgement (ACK). We assume a dense deployment of small cells where some Wi-Fi APs operate on the same channel and others on a different channel.
We proceed to evaluate the issue of association by dynamic CSAT for the LTE-U transmission and validate the effect on Wi-Fi beacon transmission and reception.

%Fig.~\ref{asso} (a) shows the deployment configuration being considered with both Cell A and Cell B using Wi-Fi on the same channel, where the users associated with Cell A and Cell B are denoted by red and blue respectively and Cell C and Cell D configured with different channel, where the users associated with Cell C and Cell D are denoted by black and brown respectively. Fig.~\ref{asso} (b) shows Cell A switching to LTE-U. We proceed to evaluate the issue of association by dynamic CSAT for the LTE-U transmission and validate the effect on Wi-Fi beacon transmission and reception.

\begin{table}[htb!]
%\label{table2}
\caption{Beacons Transmission Parameters}
\vspace{-0.2cm}
\centering
\begin{tabular}{|c|c|}
\hline
\bfseries Parameter &\bfseries Value \\ [0.4ex]
\hline\hline
DIFS, SIFS, CWmin ($W$) & 34 ($\mu s$), 16 ($\mu s$), 16 \\ 
\hline
Beacon Length ($N_b$) & 305 bytes \\
\hline
Data Rate ($r_0$), Time slot ($t_s$) & 6 Mbps, 9 ($\mu s$) \\
\hline
Phy Preamble ($t_{pr}$) &  20 ($\mu s$)  \\
\hline
Beacon transmission time ($T_b$) & 427 ($\mu s$) \\
\hline
Beacon transmission time slots ($T_{tb}$) & 48 \\
\hline
Acknowledgment (ACK) & 72 ($\mu s$) \\
\hline
Beacon transmission interval ($B_{int}$) & 102.4 ms \\
\hline
Unicast transmission time ($T_u$) & Unicast size ($N_u$) \\
\hline
\end{tabular}
\label{tab: beaconpara}
\vspace{-0.2cm}
\end{table} 

\begin{figure}[!htb]
  \includegraphics[width=9cm,height=3cm]{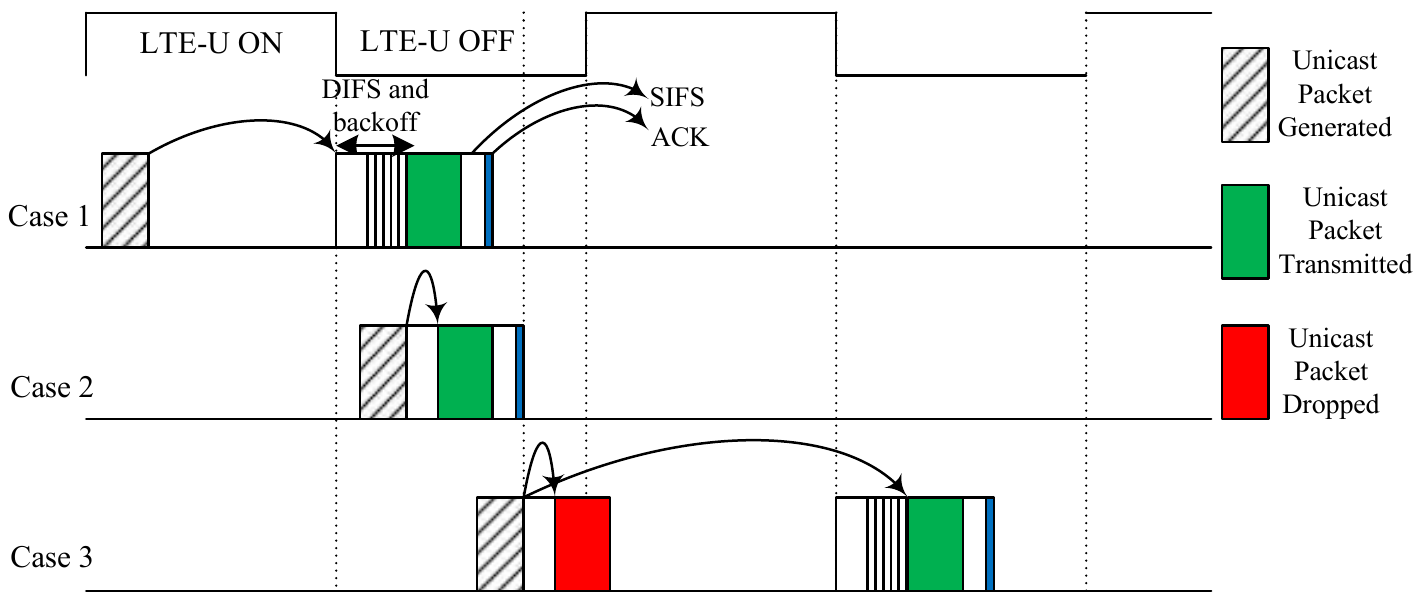}
   
  \caption{Possible cases for unicast packet transmission for the active scanning and association}
  \label{fig: unicastdrop}
  \vspace{-0.3cm}
\end{figure}
\vspace{-0.1cm}
\subsection{Unicast Packet for Association and Active Scanning}
\label{p32}

In Wi-Fi association, the probe response in active scanning, association request, association response, authentication request, and authentication response packets in both active and passive scanning are all unicast packets where the packet transmission is followed by an ACK packet. This process, in the presence of coexisting LTE-U, is explained in Fig.~\ref{fig: unicastdrop}. In Case 1, the unicast packet is generated during the ON period and, depending on the type of unicast packet, it may be generated at the Wi-Fi AP or a Wi-Fi client. In this case, the Wi-Fi node waits until the end of the ON period, senses the channel for a time equal to DIFS, selects a random back-off and transmits the packet if the channel is idle. The minimum length of the OFF period is 1 ms which is ``usually'' larger than the unicast packet transmission time which is $T_u=t_{pr}+N_u\times 8/r_0+\text{SIFS}+\text{ACK}$ as given in Table~\ref{tab: beaconpara} plus the DIFS and back-off time, so overlap with the second ON period does not happen in this case and the packet is transmitted and received successfully (assuming that there is just one of these packets in the channel, i.e. one Wi-Fi station performs the association with AP, so there is no collision between the packets. This is also assumed for Case 2 and 3). 

In Case 2, the unicast packet is generated during the LTE-U OFF period: since the channel is idle, the node performs DIFS and transmits. In this case also the unicast packet transmission time ($T_u$) plus DIFS is such that the beacon does not overlap with the second ON period and the beacon is transmitted and received successfully. 
In Case 3, the unicast packet is generated in the LTE-U OFF period and the DIFS sensing shows that the channel is idle. Two situations may happen in this case: (a) the channel is idle for DIFS and the node transmits the unicast packet but the packet or it's ACK overlaps with the second ON period, in this case the nodes retransmit in the second ON period with DIFS sensing and back-off, or (b) before finishing the DIFS sensing the second ON period starts which causes the node to wait till the end of the LTE-U ON period, then sense for DIFS and back-off for transmission. 

The presence of unicast packets in the channel makes the correct beacon transmission and reception more difficult during the LTE-U OFF duration because part of the OFF period is used by the unicast packets or the unicast packets may collide with beacon packets.
\vspace{-0.3cm}
\begin{figure}[!htb]
  \includegraphics[width=9cm,height=5.3cm]{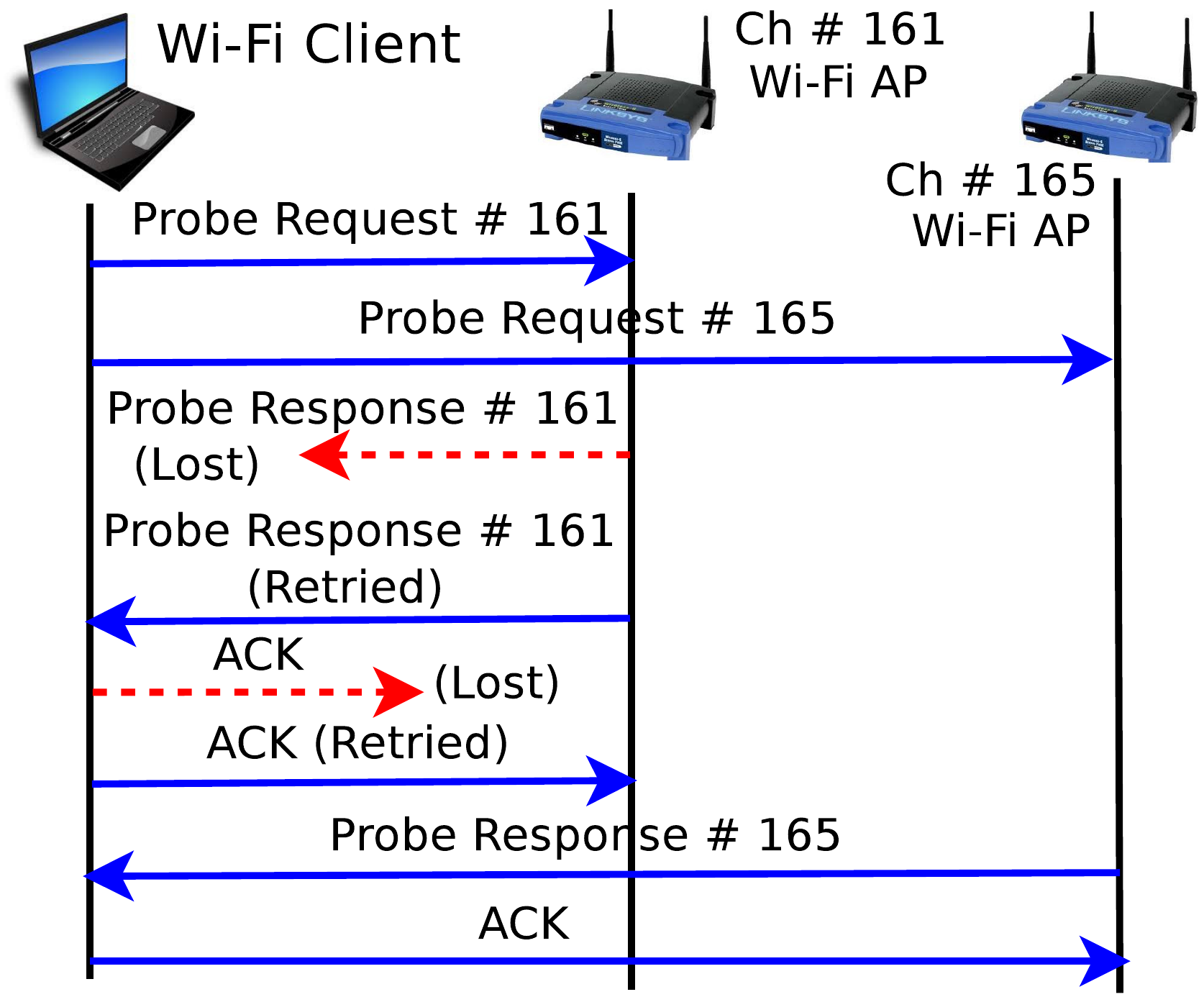}
   \vspace{-0.5cm}
  \caption{Exchange of Probe packets}
  \label{fig: act}
\end{figure}
%\vspace{-0.1cm}
\section{Motivation for 80\% CSAT Duty Cycle}\label{motiv}

As per the LTE-U forum specification,  the LTE-U BS  starts  with a  50\%  duty cycle mode and observes the medium for every N (e.g. N = 30) LTE-U OFF durations (which is equivalent to the reception of five Wi-Fi beacons). If the energy detection value during this period is greater than the threshold, the CSAT algorithm detects the presence of  Wi-Fi and maintains the same 50\% duty cycle, otherwise the LTE-U BS switches to a 95\% duty cycle. Fig.~\ref{fig: act} shows the re-transmission process when Wi-Fi probe response packets are lost. These packets are sent during active scanning for association. Interestingly, since the energy-based CSAT algorithm just detects energy levels in the medium, it will also detect energy due to the presence of probe packets, which may help it to detect the presence of Wi-Fi faster and scale back the duty cycle. However, the probe packets may cause dropped Wi-Fi beacons causing the Wi-Fi association process itself to be delayed. 
\begin{figure}[htb!]
\begin{center}
\includegraphics[totalheight=3.2cm,width=7.8cm]{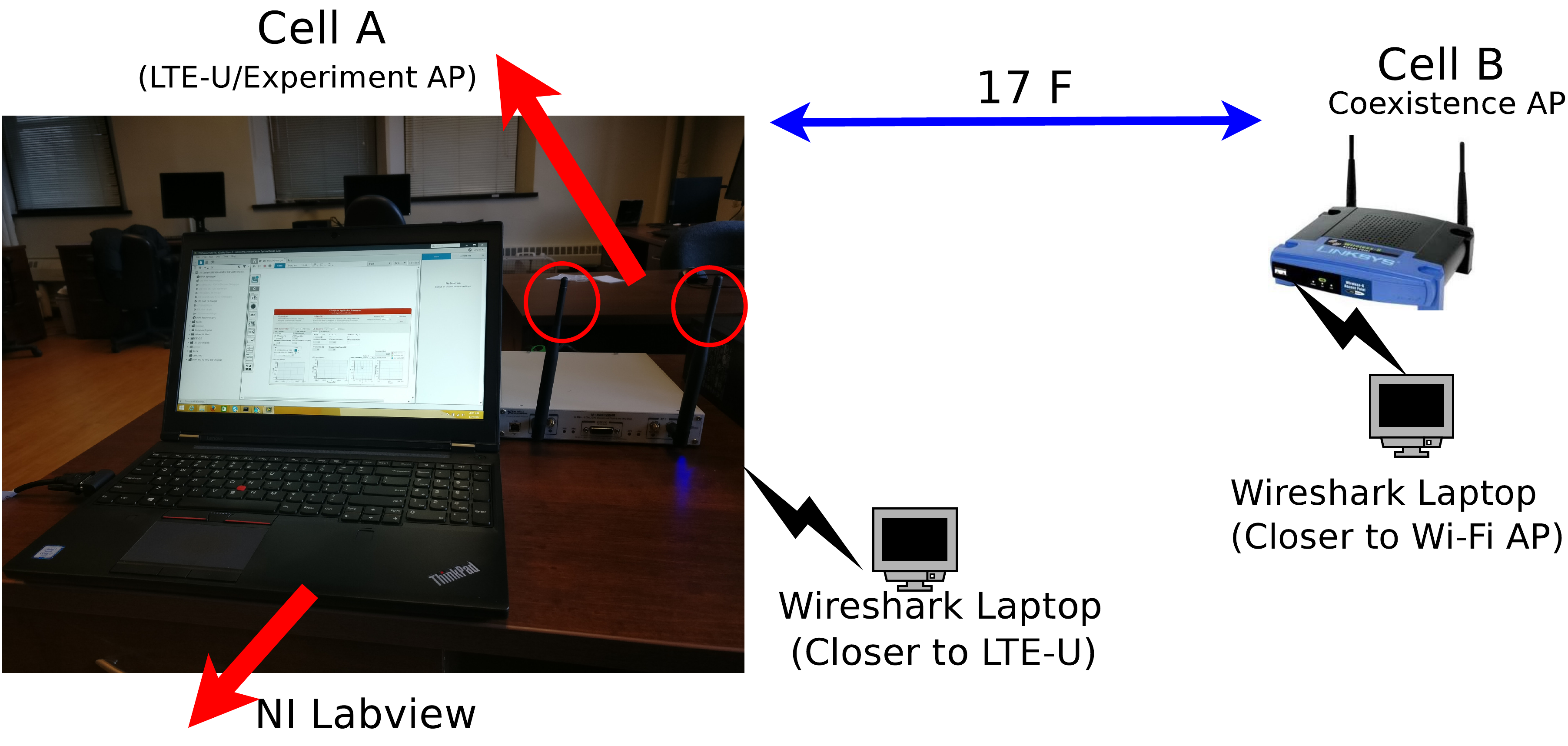}
\vspace{-0.1cm}
\caption{LTE Wi-Fi Co-existence Experimental Setup.}
 \vspace{-0.5cm}
\label{tbed}
\end{center}
\end{figure}

In order to study the impact of dropped beacons when LTE-U operates at 95\% duty cycle, we set up an experimental test-bed using the NI USRP platform as shown in Fig. ~\ref{tbed}, where Cell A is a LTE-U BS and Cell B is a Wi-Fi AP. Cell B initially does not transmit any data, only beacon frames (and probe responses, if clients in the vicinity transmit probe requests) as explained in more detail in Section VI. We ensure that both Wi-Fi AP and LTE-U BS have the same system time reference. When the Wi-Fi AP is switched ON, it starts the first beacon transmission at 54.112 seconds (observed from Wireshark’s packet trace on a laptop which is placed very close to the Wi-Fi AP) and the LTE-U BS detects the Wi-Fi AP beacon packet at 54.591 seconds. Thus it is clear that some initial beacon packets were not detected by the LTE-U BS.
\vspace{-0.4cm}
%\vspace{-0.4cm}
\begin{figure}[!htb]
 \centering
  \includegraphics[width=9cm,height=5.3cm]{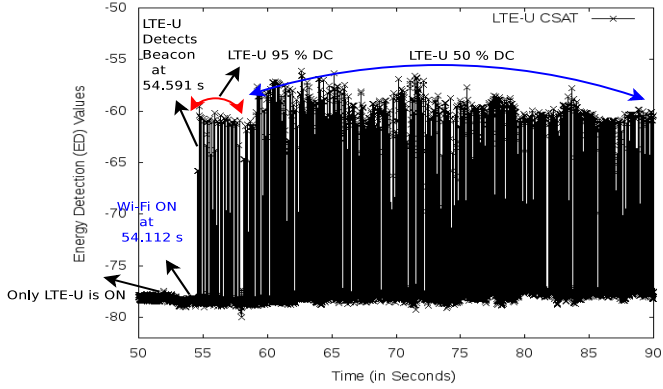}
 \vspace{-0.4cm}
  \caption{Energy Based CSAT Algorithm at 95\% Duty Cycle}
  \label{eb}
\end{figure}
%\vspace{-0.3cm}

Fig.~\ref{eb} clearly shows the operation of the energy-based CSAT algorithm when LTE-U starts with a 95\% duty cycle and then a Wi-Fi AP turns on in the same channel and begins the association process. The presence of probe request, probe response, authentication request, authentication response, association request, association response, ACK and beacon packets, combined with the short 1 ms OFF duration, causes a lot of beacons to be dropped initially and thus causes the CSAT algorithm to take more time to scale back the duty cycle. The decision making (or) measuring time for LTE-U BS after observing the first K (for eg., k = 5) beacon packets is roughly 55.520 seconds, this delay is due to the drop in initial beacons (because of probe and association packets). The other additional delay due to the NI hardware is 58.730 seconds. Hence, the total time for the energy based CSAT algorithm to adopt or change the duty cycle is 4.618 seconds (\emph{i.e.,} Wi-Fi 1st beacon transmission time, 54.112 Seconds + LTE-U detects $K$ beacon packets time, 55.520 Seconds + NI hardware processing time, 58.730 Seconds). The successful reception percentage of beacon at 95\% duty cycle CSAT is 79.3\% (refer Fig.~\ref{breception} in Section VI). This observation motivates the use of a 80\% duty cycle instead of a 95\% duty cycle in order to increase the beacon reception success rate and reduce the scale back time.

\section{Energy Based CSAT Algorithm}

The ability of the the LTE-U BS to detect the Wi-Fi energy level present on the current channel depends on the noise floor, ambient energy and interference sources. In addition, energy detection requires a \textit{pre-defined threshold} which determines if the reported energy level is adequate to report the transmission medium as busy or idle.
\begin{algorithm}
\vspace{-0.1cm}
 \caption{Energy Based CSAT Beacon Transmission}
 \begin{algorithmic}
 \item \textit{Initialization:} $(i)$ LTE-U operates at 50\% duty cycle,\\ 
\hspace{2cm}$(ii)$ Count 1 = 0, Count 2 = 0;
% \end{itemize}
 \While {true}         
     /*  Dynamic LTE-U Duty Cycle. */
     \For {k = 1 to N}  \\
     /*  Observe the medium for N (\emph{e.g.,} 30 for 20 ms ON/OFF cycle) such LTE-U OFF duration. */
       \If {Energy Detection ($ED$) $\geq$ Threshold}\\
       /* We assume the ED threshold is $-70$ dBm */ \\
    /* If ED is greater than threshold then there is a possible of Wi-Fi beacons or Probe request packets or Probe response packets got detected. */
        \State Count 1 ++; 
     \Else \\
     /* No Wi-Fi beacons got detected. */
     \State Count 2 ++; 
         \EndIf
     \EndFor
     \If {((Count 1 $\geq$ 5) \&\& (Avg(ED) $\geq$ Threshold))} \\ /* Observe first five Wi-Fi beacon packets. */
        \State LTE-U maintain or switches to 50\% duty cycle;
        \State Count 1 = 0;
     \Else
     \State LTE-U maintain or switches to 80\% duty cycle;
     \State Count 2 = 0; 
         \EndIf
   \EndWhile
  \end{algorithmic}
 \end{algorithm}
%\vspace{-0.2cm}
As per the LTE-U forum specification, the maximum and minimum ON duration is 20 ms and 1 ms, respectively. The minimum LTE-U OFF duration is 1 ms but in our work we kept it as 5 ms (\emph{i.e.,} a 80\% duty cycle with 20 ms ON and 5 ms OFF). The LTE-U BS chooses a duty cycle based on the CSAT procedure described in Algorithm 1. Initially, the LTE-U BS starts with a 50\% duty cycle mode and observes the medium for every 30 LTE-U OFF duration's (which is equivalent in time to five consecutive Wi-Fi beacon packets). If the energy detection value on the current channel is greater than the threshold (set at - 70 dBm), the LTE-U BS detects the presence of Wi-Fi (in terms of beacons) and increments the ``Count 1'' value. To make sure that the presence of Wi-Fi AP is detected (and not just probe request packets from Wi-Fi clients), the LTE-U BS will wait until the ``Count 1'' value is greater than or equal to five and the average of energy detection value is greater than the threshold. If these conditions are satisfied, the LTE-U BS will continue to operate with a 50\% duty cycle, otherwise the LTE-U BS will operate at 80\% duty cycle. 
Probe request and response packets help the CSAT algorithm to make the decision faster. This is due to the fact that beacon packets are scheduled by the Wi-Fi AP every 102.4 ms, but the probe requests are randomly generated by Wi-Fi clients in the vicinity and in a dense environment there are likely a large number of such clients. As the CSAT algorithm is based on energy detection alone, the algorithm will be able to capture the change in energy levels due to probe packets.
\vspace{-0.1cm}
\section{LTE-U Expected Delay for Detecting an Active Wi-Fi Link}
\label{sec: ExpectedDelay}
In this section, we theoretically analyze the expected delay for reception of $K$ consecutive Wi-Fi beacons at the LTE-U BS.
In order for LTE-U to detect and switch from a high duty cycle state (which assumes there is no Wi-Fi in the network) to a 50\% duty cycle, the CSAT algorithm has to detect the energy of several beacon packets to make sure that Wi-Fi is active. In the theoretical derivation, we neglect the presence of probe request and response packets and assume that the Wi-Fi AP performs only passive scanning for association. So, the LTE-U needs to detect the energy of $K$ consecutive beacons to identify the presence of Wi-Fi. The expected delay directly depends on the beacon drop probability; as the beacon drop probability increases the LTE-U expected delay to detect an active Wi-Fi link increases. Thus, we first calculate the beacon drop probability and then the expected delay. 

For the beacon drop probability calculation, we assume that the Wi-Fi and LTE-U are co-channel and all Wi-Fi stations (AP and clients) can perfectly detect the LTE-U in the ON period. The beacon transmission follows the CSMA/CA protocol and since there is no ACK, the random back-off is selected based on the minimum contention window ($CW_{min}$). The overlap of Wi-Fi beacon with the LTE-U ON period (as the LTE-U in not sensing for transmission) is the reason for the beacon drop. This is explained in detail in Section III 3a in \cite{WCNC}). We assume that any partial overlap of LTE-U with Wi-Fi beacon results in the beacon packet collision. The probability that a beacon is generated at a time slot in the OFF period is:
\begin{equation}
\begin{split}
P_t &= P(OFF)P(t_s|OFF)=\frac{T_{OFF}}{T_{ON}+T_{OFF}}\times \frac{t_{s}}{T_{OFF}}\\
&=\frac{t_{s}}{T_{ON}+T_{OFF}},
\label{ed: Pt}
\end{split}
\end{equation}
where $T_{ON}$ is the LTE-U ON, and $T_{OFF}$ is LTE-U OFF period. So, the beacon drop probability given the beacon airtime ($T_b$) is calculated as:
\begin{equation}
\begin{split}
P_{d} &=P_t(\lceil T_b/t_s\rceil).
\end{split}
\label{ed: Pd}
\end{equation}
From eq (\ref{ed: Pd}), we see that the drop probability is independent of the LTE-U duty cycle and depends only on the total time of one ON/OFF cycle, \emph{i.e.,} it depends on the number of ON/OFF cycles which happen in one beacon transmission period $T_d=102.4$ ms (or equivalently the number of ON edges in a 102.4 ms period). The probability of time interval between the successful detection of the beacons by the LTE-U (i.e. the interval between the transmitted beacons which are not colliding with the LTE-U ON edge) follows the geometric distribution. Let the $s_k$ be the time interval between the successful detected beacons, so the probability can be calculated as:
\begin{equation}
\begin{split}
P_{s}(k) =P(s_k=iT_d)=(1-P_d)P_d^{(i-1)},
\end{split}
\label{ed: Ps}
\end{equation}
Thus, the expected time interval for a beacon to be successfully detected can be calculated as:
\begin{equation}
\begin{split}
E[s_k]&=\sum_{i=0}^{i=\infty} i T_d (1-P_d)P_d^{(i-1)}\\
& =T_d (1-P_d)  \frac{1}{(1-P_d)^2}=\frac{T_d}{1-P_d}
\end{split}
\label{ed: Es}
\end{equation}
The total expected delay for the LTE-U to detect $K$ successful beacon for detecting an active Wi-Fi link is calculated as:
\begin{equation}
\begin{split}
D_l&=\sum_{k=1}^{k=K} E[s_k]=K\frac{T_d}{1-P_d}
\end{split}
\label{ed: Dl}
\end{equation}
\begin{table}[htb!]
\label{table2}
\caption{Simulation parameters}
\vspace{-0.1cm}
\centering
\begin{tabular}{|p{4cm}| p{4cm}|}
\hline\bfseries
Parameter&\bfseries Value \\ [0.4ex]
\hline
Bandwidth & 20 MHz  \\
\hline
Operating frequency & 5.805 GHz (\emph{i.e.,} Channel 161)  \\
\hline
Tx power of LTE and Wi-Fi & 23 dBm \\ 
\hline
Wi-Fi Energy Threshold & -82 dBm \\
\hline
Wi-Fi and LTE-U Antenna Type & MIMO \& SISO\\
\hline
LTE-U data and control channel & PDCCH and PDSCH \\
\hline
\end{tabular}
\label{table:Building1}
 %\vspace{-0.5cm}
\end{table} 

\begin{figure*}[!htb]
\begin{subfigure}[c]{0.5\textwidth}
  \centering
  \includegraphics[height=5cm,width=9cm]{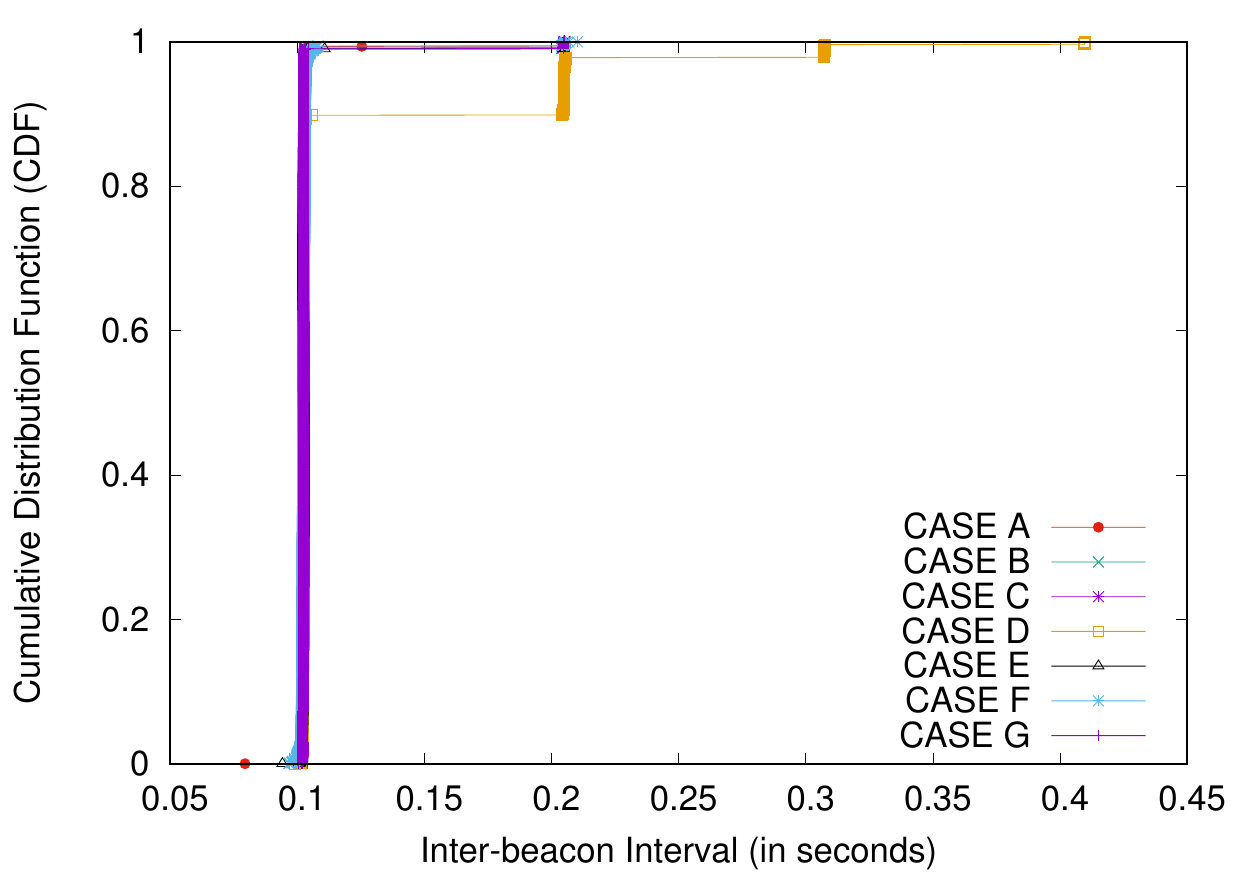}
\end{subfigure}
\begin{subfigure}[c]{0.5\textwidth}
  \centering
  \includegraphics[height=5cm,width=9cm]{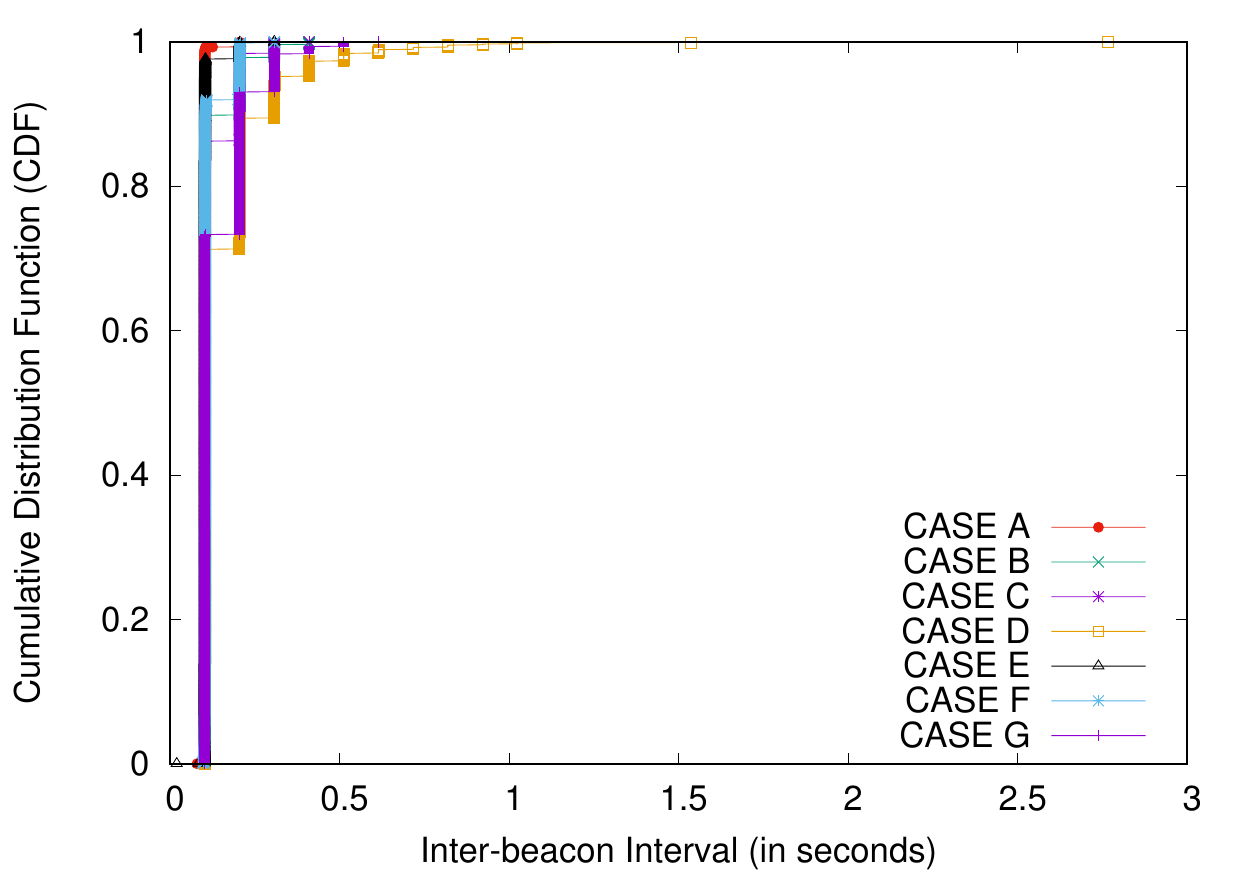} 
\end{subfigure}
\caption{(a) CDF of transmitted beacon intervals  and (b) CDF of received beacon intervals}
\vspace{-0.6cm}
 \label{be}
\end{figure*}

 \section{EXPERIMENTAL RESULTS AND ANALYSIS}
In this section, we first describe the experimental setup, followed by comparison of LTE-U using adaptive CSAT with different LTE-U fixed duty cycles at both Wi-Fi \& LTE-U reception side. We also validate the theoretical and experimental results for K successful beacon reception time.

\subsection{Experimental Setup}
In Section II, the co-existence system model stated uses a setup consisting of a NI USRP, off-the-shelf Wi-Fi APs and other client devices. An open-lab environment was created for the experiment, where there are other Wi-Fi clients in the area transmitting probe requests to the APs being tested. The LTE coexistence framework is deployed and we configure the NI USRP 2953R SDR to transmit a LTE-U signal. To obtain different duty cycles, we vary the ON and OFF cycle times. The Wi-Fi APs stated earlier are two Netgear APs. Fig.~\ref{tbed} shows the experimental test-bed setup. Cell A could be either the LTE-U BS or a Wi-Fi AP (labeled as Experiment AP). Cell B is always configured to be a Wi-Fi AP labelled as Coexistence AP, and does not transmit any data but only beacon frames (and probe requests, if any clients in the vicinity transmit probe requests). Both the Wi-Fi APs and the NI LTE-U BS are provisioned to operate on the same unlicensed channel (Channel 161). We also made sure there are no other Wi-Fi APs on this channel. The NI LTE-U BS implemented the CSAT algorithm described in Section IV. We measure how many Wi-Fi beacons are received at the LTE-U BS (Cell A) by using Wireshark in monitor mode on a laptop placed very close to the LTE-U BS. Beacons received by this laptop are labeled as ``Wi-Fi beacons received at Cell A". Similarly, we placed a monitoring device close to the Wi-Fi AP (Cell B), and the beacons received by this laptop are labeled as ``Wi-Fi beacons received at Cell B". We compare how many beacons were successfully transmitted between the laptop and their respective cells by checking the sequence IDs of each beacon that is captured by Wireshark at each laptop. These characteristics are tabulated in Table~\ref{table:Building1}.

\vspace{-0.1cm}
\subsection{Beacon Detection for different LTE-U duty cycles}
In this section, we examine the beacon reception performance for different LTE-U duty cycles and energy based CSAT algorithm with realistic Wi-Fi AP configuration of both active and passive scanning. We will consider the following cases: \textbf{Case A:} Wi-Fi/Wi-Fi Coexistence, \textbf{Case B:} LTE-U/Wi-Fi Coexistence with fixed 5 ms ON/5 ms OFF LTE-U duty cycle, \textbf{Case C:} LTE-U/Wi-Fi Coexistence with fixed 20 ms ON/20 ms OFF LTE-U duty cycle, \textbf{Case D:} LTE-U/Wi-Fi Coexistence with fixed 20 ms ON/1 ms OFF LTE-U duty cycle, \textbf{Case E:} LTE-U/Wi-Fi Coexistence with fixed 20 ms ON/5 ms OFF LTE-U duty cycle,
\textbf{Case F:} LTE-U operates at 80\% duty cycle when there is no Wi-Fi and operates at 50\% duty cycle when CSAT detects Wi-Fi, \textbf{Case G:} LTE-U operates at 95\% duty cycle when there is no Wi-Fi and operates at 50\% duty cycle when CSAT detects Wi-Fi.

%\vspace{-0.1cm}
\begin{figure}
\begin{center}
\includegraphics[height=4.8cm,width=9cm]{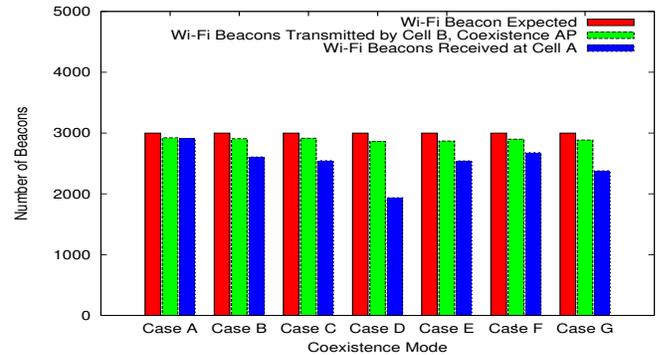}
\vspace{-0.5cm}
\caption{Beacons Reception at different ON/OFF modes.}
\label{breception}
\end{center}
\vspace{-0.7cm}
\end{figure}
In the cases stated above, all the measurements for beacon transmissions, receptions and inter-beacon intervals were over a time period spanning 5 minutes. In these 5 minutes, it is expected that 3000 beacons would be transmitted. These beacons are transmitted by the Coexistence AP in Cell B to the LTE-U BS in Cell A, which is 17 feet away from Cell B as depicted in Fig.~\ref{tbed}. The CDF of transmitted beacon and received beacon intervals are shown in Figs.~\ref{be} (a) and~\ref{be} (b). We see that the transmitted and received beacon interval increases in Case D and Case G considerably, compared to the other scenarios, with the received beacon interval extending to seconds in some cases.

The total number of expected, transmitted and received beacons for each of the seven cases are shown in Fig.~\ref{breception}. In Case A \emph{i.e.,} during Wi-Fi/Wi-Fi co-existence, there is no notable drop in the number of beacons received. In all of the other cases of LTE-U/Wi-Fi co-existence however, there is a noticeable drop in the number of beacons received, even when the duty cycle is 50\% (represented by Case B and Case C). In Case D which is 20 ms ON and 1 ms OFF, almost one-third of the transmitted beacon frames are not received at the LTE-U BS. An interesting observation is that in Case E which is 80\% duty cycle, the beacon loss is comparable to the Case B and Case C which were at 50\% duty cycle. From this, we conclude that it may be advisable for a LTE-U BS to use a duty cycle of at most 80\% to allow a fair access for the Wi-Fi AP in the medium. With 80\% duty cycle, the reduction in the number of beacons that are not received at the LTE-BS will also enable the LTE-U BS to react faster to the presence of Wi-Fi by reducing its duty cycle to the required 50\%. 
\par In order to validate the above claim we compared the expected beacon transmission and reception with 80\% duty cycle CSAT (\emph{i.e.,} Case F) and 95\% duty cycle CSAT (\emph{i.e.,} Case G). The 80\% CSAT performs better with respect to beacon reception compared to the 95\% CSAT and it achieves performance close to the Wi-Fi/Wi-Fi coexistence (\emph{i.e.,} Case A). In 80\% duty cycle CSAT most of the beacon transmission is successful because all the probe, association and ACK can get through easily with out any overlap in the 5 ms LTE-U OFF duration. In order to achieve 90\% successful beacon reception, the LTE-U BS should operate in Case F mode.  

 \begin{figure}[htb!]
 \vspace{-0.2cm}
\begin{center}

% % \epsfig{width=9cm,figure=table.eps}
\includegraphics[height=5cm,width=9cm]{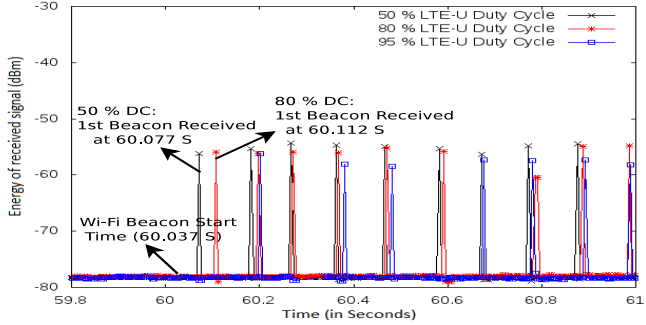}
\vspace{-1.2cm}
\caption{Beacon reception time at LTE-U side}
\label{wifi}
\vspace{-0.4cm}
%\subcaption{Aerial view of floor area inside the building.}
\label{br}
\end{center}
\end{figure}

Fig.~\ref{br} shows the beacon reception time at LTE-U side over different duty cycles. In this experiment, we ensure that the same start time for the different duty cycle is maintained by ensuring the same system time (in terms of minutes and seconds). Hence, these three experiments (\emph{i.e.,} 50\% (represented by black), 80\% (represented by red), 95\% duty cycle (represented by blue)) start at the same time, \emph{i.e.,} at 60.37 seconds. The first beacon packet received at 80\% duty cycle is faster compare to 95\% duty cycle; also, we observe that more beacons drop at 95\% in a short time scale. The total time for the 80\% duty cycle CSAT algorithm to adopt is 2.980 seconds (\emph{i.e.,} Wi-Fi 1st beacon transmission time + LTE-U detects \textit{K} beacon packets time + NI hardware processing time) and for 95\% CSAT is 4.618 seconds (refer Fig.~\ref{eb} in section \ref{motiv}). Hence, the overall scale back time for 80\% is faster compared with the 95\% duty cycle CSAT.
\begin{table}[!htbp]
\caption{The expected delay for detecting $K=5$ successful beacon to detect an active Wi-Fi link.}
\vspace{-0.3cm}
\label{table: Del}
\begin{center}
\begin{tabular}{|c|c|c|}
\hline
Setup & Theoretical & Experimental \\ 
\hline\hline
$T_{ON}=T_{OFF}=$ 5ms & 535.62 ms& 531.89 ms \\ 
\hline
$T_{ON}=$ 20ms, $T_{OFF}=$ 1ms & 522.76 ms & 529.94 ms \\ 
\hline
$T_{ON}= 20 ms, T_{OFF}=$ 5ms & 521 ms &  528.1 ms \\ 
\hline
\end{tabular}
\end{center}
\vspace{-0.4cm}
\end{table}

\subsection{Expected Delay for \textit{k} Successful Beacon Reception}

To validate the theoretical derivation of expected delay for receiving \textit{K} successful beacon as explained in Section~\ref{sec: ExpectedDelay} we compared it with the experimental results. Since the theoretical derivation assumed no packets other than beacon packets, we needed to perform the experiment under conditions where no Wi-Fi clients were present to transmit probe requests. Hence, this experiment was performed late at night on the University of Chicago campus and we used Wireshark to confirm that indeed only beacon packets were being transmitted by the Wi-Fi AP and there were no other packets on the air. Under these conditions, we see from Table~\ref{table: Del} that there is very good agreement between the theoretical expected delay for the \textit{k} = 5 successful beacon as calculated from eq~\eqref{ed: Dl} and the experimental values from the NI test-bed.

We also observe from Table~\ref{table: Del} that there does not seem to be much difference between 50\%, 80\% and 95\% duty cycles. This is because we have neglected the impact of probe request and response packets, whose effects are difficult to model analytically. In practice, probe request and response packets will always be present in the channel along with the beacons. This will lead to an increase in the beacon drop probability and correspondingly the expected delay increases as indicated in eq~\eqref{ed: Dl}. We would then expect to see a larger expected delay difference between the three scenarios in Table~\ref{table: Del} which indicates that the choice of duty cycle is very critical for fast association. We see this from the experimental results in Fig.~\ref{breception} (with active and passive scanning modes where probe packets and beacon packets are being transmitted) where the 95\% duty cycle (Case G) has more beacon drops compared to 80\% (Case F). Thus, in realistic conditions the gap between 95\% and 80\% duty cycle in Table~\ref{table: Del} will be larger. Therefore, in order to scale back our CSAT algorithm faster, the LTE-U should operate at 80\% duty cycle when there is no Wi-Fi AP.  

\section{Conclusions and Future Work}\label{p32}

In this paper, we carefully analyzed, theoretically and experimentally, the performance of an energy based CSAT algorithm that detects Wi-Fi and scales back the LTE-U duty cycle automatically, in particular the behavior during the Wi-Fi association process. The motivation was to understand if LTE-U should be transmitting with its maximum allowed duty cycle of 95\% when it is operating on an empty channel. We find that if it does so, it will severely impact the ability of a Wi-Fi AP to begin sharing the channel since the beacon transmission/reception will be disrupted, i.e. the Wi-Fi stations cannot associate in a fair amount of time. Instead, if LTE-U scaled back the occupancy on an empty channel to 80\% CSAT (i.e. 20 ms ON/5 ms OFF), the Wi-Fi beacon loss reduces to an acceptable level and makes it easier for Wi-Fi to get on the air.\\
This work was supported by NSF under grant CNS-1618920.

\end{document}